\documentclass[3p,times]{elsarticle}

\usepackage{ecrc}

\usepackage{epstopdf}
\volume{00}

\firstpage{1}

\journalname{Nuclear Physics A}

\runauth{Author1 et al.}

\jid{nupha}

\jnltitlelogo{Nuclear Physics A}

\usepackage{graphicx}
\usepackage{amsmath,amssymb}
\usepackage{float}
\usepackage{subfigure}
\begin{document}

\begin{frontmatter}

%% Title, authors and addresses

%% use the tnoteref command within \title for footnotes;
%% use the tnotetext command for the associated footnote;
%% use the fnref command within \author or \address for footnotes;
%% use the fntext command for the associated footnote;
%% use the corref command within \author for corresponding author footnotes;
%% use the cortext command for the associated footnote;
%% use the ead command for the email address,
%% and the form \ead[url] for the home page:
%%
%% \title{Title\tnoteref{label1}}
%% \tnotetext[label1]{}
%% \author{Name\corref{cor1}\fnref{label2}}
%% \ead{email address}
%% \ead[url]{home page}
%% \fntext[label2]{}
%% \cortext[cor1]{}
%% \address{Address\fnref{label3}}
%% \fntext[label3]{}

\title{Measurement of heavy-flavour production as a function of multiplicity in pp and p--Pb collisions with ALICE}

%% Single author (and collaboration) - please insert
\author{Riccardo Russo (for the ALICE Collaboration)}

\address{Dipartimento di Fisica, Universita' degli Studi di Torino and INFN, via Pietro Giuria 1, 10125 Torino}

%% For multiple authors, replace the above by:

%\author[label1]{Author1}
%\author[label2]{Author2}

%\address[label1]{Address 1}
%\address[label2]{Address 2}

\begin{abstract}
%% Text of abstract

In these proceedings results are presented from the measurement of open heavy-flavour production as a function of charged-particle multiplicity in pp collisions at  $\sqrt {s}$ = 7 TeV  and p--Pb collisions at  $\sqrt {s_{\rm NN}}$=5.02 TeV recorded with the ALICE detector in 2010 and 2013, respectively.  $\rm D^{0}$, $\rm D^{+}$ and $\rm D^{*+}$ mesons are reconstructed from their hadronic decay channels in the central rapidity region, and their production yields are measured in various multiplicity and $p_{\rm T}$ intervals.  

The per-event yields of $\rm D$ mesons in the various multiplicity intervals, normalized to their multiplicity-integrated value, and their evolution with $p_{\rm T}$ are measured for pp and p--Pb collisions to study the contribution of 
Multi-Parton Interactions (MPIs) to open charm production in the two systems. 
The nuclear modification factor of $\rm D$ mesons in p--Pb collisions, defined as the ratio of the $\rm D$-meson yields in p--Pb and pp collisions scaled by the average number of binary collisions $\langle N_{\rm coll}\rangle$, is discussed in terms of its dependence on the event activity.
Several experimental estimators
of the event activity are used in order to assess the role of kinematic
biases.
\end{abstract}

\begin{keyword}
%% keywords here, in the form: keyword \sep keyword
QGP \sep Heavy-flavour \sep Multi-Parton Interaction
%% MSC codes here, in the form: \MSC code \sep code
%% or \MSC[2008] code \sep code (2000 is the default)

\end{keyword}

\end{frontmatter}

%%
%% Start line numbering here if you want
%%
% \linenumbers

%% main text

\section{Introduction}
\vspace{-0.2cm}
\label{intro}
The measurement of heavy-flavour production as a function of the
multiplicity of charged particles produced in hadronic collisions is sensitive to
the interplay between hard and soft contributions to particle production and  could give insight into the role of Multi-Parton Interactions
(MPIs, i.e. several hard partonic interactions occuring in a single collision
between two nucleons). 

Particle production at the LHC is expected to have a substantial contribution from MPIs in pp (p--Pb) collisions, where the highest multiplicity values observed
are similar to the ones of peripheral Cu--Cu (Pb--Pb) collisions at RHIC (LHC). 
Measurements by the CMS Collaboration  of jet and underlying event properties have shown better agreement with models including MPIs \cite{CMSMPI}.  Measurements by the ALICE
Collaboration of minijets point to an 
increase of MPIs with increasing  charged-particle multiplicity \cite{Minijets}. 
In the heavy-flavour sector several measurements have been performed  \cite{NA27,LHCb}. In particular ALICE found an approximately linear increase of J/$\psi $ yield as a function of multiplicity in pp 
collisions at $\sqrt {s}$ = 7 TeV \cite{JPsiMult}. 

Moreover it is interesting to compare heavy-flavour production in p--Pb collisions with pp results to test
whether the yield and the transverse momentum distributions follow a scaling with the number of binary
nucleon--nucleon collisions in the p--Pb collision. This scaling is expected for particles
produced in hard (high virtuality) partonic scattering processes in the absence of nuclear effects in the initial or in the final
state of the p--Pb collision.
This is studied by measuring the nuclear
modification factor  $R_{\rm pPb}$,  defined as the ratio of the $p_{\rm T}$-differential cross
section measured in p--Pb collisions to that measured in pp collisions scaled by the mass
number A of the Pb nucleus.
The D-meson $R_{\rm pPb}$ in minimum bias p--Pb collisions was found  consistent with unity for $p_{\rm T}>$ 1 GeV/$c$
within uncertainties of about 20\% \cite{RpPb}, showing that  Cold Nuclear Matter (CNM) effects (nuclear modifications of the parton distribution functions (PDF), $k_{\rm T}$ broadening, energy loss in cold nuclear 
matter) do not strongly affect charm-quark production in p--Pb collisions \cite{Shuang}.  It is interesting to
measure the nuclear modification factor in p--Pb collisions in  classes of the event
activity, because the latter is related to the collision centrality (i.e. impact parameter as well as the number of participating nucleons
and  binary collisions) of the p--Pb collisions. The measurement of the $R_{\rm pPb}$ requires to estimate the average number of binary collisions $\langle N_{\rm coll}\rangle$ for the
event activity  intervals used in the analysis.  ALICE has identified in p--Pb collisions several
sources that can induce a bias in the centrality determination based on
particle multiplicity measurement  \cite{Andreas, Alberica}. This bias have been  observed in the measurement of the
nuclear modification factor of charged particles in p--Pb collisions in
multiplicity classes \cite{Alberica}. This work investigates whether  such a bias is also present for $\rm D$ mesons.

The results are presented in form of the $\rm D$-meson self-normalized yield  in pp and p--Pb collisions, defined as
\vspace{-0.2cm}
\begin{equation}
\frac{\rm d^2\textit{N}^{\rm D}/d\textit{y}d\textit{p}_{\rm T}}{\langle \rm d^2\textit{N}^{\rm D}/d\textit{y}d\textit{p}_{\rm T} \rangle}=\frac{Y^{\rm mult}/(\epsilon^{\rm mult})}{Y^{\rm tot}/(\epsilon^{\rm tot}\times \epsilon^{\rm trigger})}
\end{equation}
where $Y^{\rm mult}$ is the D-meson per-event yield in multiplicity intervals, $Y^{\rm tot}$ the multiplicity integrated per-event yield, $\epsilon$ are the corresponding reconstruction and selection efficiencies and $\epsilon^{\rm trigger}$ is the trigger efficiency (only relevant in pp).  Furthermore, the
binary scaling is studied in several event activity classes via the $Q_{\rm pPb}$
ratio, defiled as:
\vspace{-0.2cm}
\begin{equation}
Q_{\rm pPb}^{\rm V0A}(p_{\rm T}) = \frac{\rm{d} \textit{N}_{mult}^{\rm pPb}/\rm d \textit{p}_{\rm T}}{\langle N^{\rm Glauber}_{\rm coll}\rangle \rm{d} \textit{N}^{\rm pp}/\rm d \textit{p}_{\rm T}} \ \ \ \ \ \ \ \ \ \ \ \ \ \ \ \ \ \ Q_{\rm pPb}^{\rm mult}(p_{\rm T}) = \frac{\rm{d} \textit{N}_{mult}^{\rm pPb}/\rm d \textit{p}_{\rm T}}{\langle N^{\rm mult}_{\rm coll}\rangle \rm{d} \textit{N}^{\rm pp}/\rm d \textit{p}_{\rm T}}
\end{equation}
for the V0A and ZNA event activity estimators respectively, defined in the next section.
 These two observables represent different
ways to study the multiplicity dependence of $\rm D$-meson production in p--Pb collisions:
 the self-normalized yields are more focused on the study of MPIs, while $Q_{\rm pPb}$ reflects the scaling  of charm production in p--Pb collisions relative to pp collisions.
\vspace{-0.5cm}
\section{Data sample and analysis strategy}
\vspace{-0.2cm}
The ALICE detector is described in \cite{ALICEDet}.
The data samples analyzed are from the 2010 pp run (300 $\cdot 10^6$ events at  $\sqrt {s}$ = 7 TeV) and the 2013 p--Pb run  (100 $\cdot 10^6$ events at $\sqrt {s_{\rm NN}}$ = 5.02 TeV).
Details on the trigger and event selections can be found in \cite{RpPb,Charmpp}.
Events are divided in event activity classes. Three event activity estimators have been used:
\begin{itemize}
\item $N_{\rm tracklets}$: number of track segments reconstructed in the Silicon Pixel Detector (SPD - two innermost layers of the Inner Tracking System - $|\eta|<0.9$ );
\item V0A: signal amplitude of the A-side VZERO scintillator  ($2.8<\eta<5.1$ - Pb going direction for p--Pb collisions);
\item ZNA: energy from nuclear fragments in the A-side Zero Degree Neutron Calorimeter  (112.5 m from interaction point - Pb going direction for p--Pb collisions).
\end{itemize}
The self-normalized yields are obtained in $N_{\rm tracklets}$ intervals, while
the $Q_{\rm pPb}$ analysis adopts V0A and ZNA multiplicity classes (0-20\%, 20-40\%, 40-60\%, 60-100\%).
The analysis is based on the reconstruction of $\rm D$ mesons in their  hadronic decay channels ($\rm D^{0}\rightarrow \rm  K^{-} \rm \pi^{+}$,
 $\rm D^{+}\rightarrow K^{-}\pi^{+} \rm \pi^{+}$,
 $\rm D^{*+}\rightarrow \rm D^{0} \rm \pi^{+}$)
  in the ALICE central barrel ($|\eta|<$0.9), exploiting the excellent vertex resolution and particle identification capabilities of the ALICE detector as described in \cite{io}. 
 $\rm D$-meson efficiency corrections in pp and p--Pb collisions  are determined with Monte Carlo simulations based on PYTHIA 6.4.21 and HIJING event generators. 
 A fraction of the total $\rm D$-meson  yield comes from the decay of B mesons. For the $Q_{\rm pPb}$ analysis this contribution  was estimated based on
FONLL pQCD calculations as described in \cite{io}, while for the self-normalized yield analysis no subtraction has been performed,
assuming that the fraction of feed-down $\rm D$ mesons does not depend on multiplicity and cancels in the ratio of Eq. 1. A deviation from this assumption has been considered to estimate  the corresponding systematic uncertainty.
The pp and p--Pb corrected yields have been used to compute the $Q_{\rm pPb}$ as in Eq. 2.
The average values of $N_{\rm coll}$ in the four V0A and ZNA event activity classes have been evaluated as follows:
\begin{itemize}
\item V0A: the $\langle N_{\rm coll}^{\rm Glauber}\rangle$ values have been obtained for each V0A
multiplicity interval with the approach used for Pb--Pb collisions, i.e. via a
fit to the V0A multiplicity distribution based on the Glauber model for the
collision geometry and a two-component model for particle
production \cite{Cent}.
\item ZNA: the $\langle N_{\rm part}^{\rm mult}\rangle$ values have been
calculated by scaling the $\langle N_{\rm part}\rangle$ in minimum-bias p--Pb collisions by the
ratio between the average multiplicity density measured at mid-rapidity for a
given ZN energy event class and the one measured in minimum bias
collisions. $\langle N_{\rm coll}^{\rm mult}\rangle$ is obtained as  $\langle N_{\rm coll}^{\rm mult}\rangle$ = $\langle N_{\rm part}^{\rm mult}\rangle$ - 1.
\end{itemize}

\vspace{-0.7cm}
\begin{figure}
 \centering
 \subfigure
   {\includegraphics[width=6.5cm]{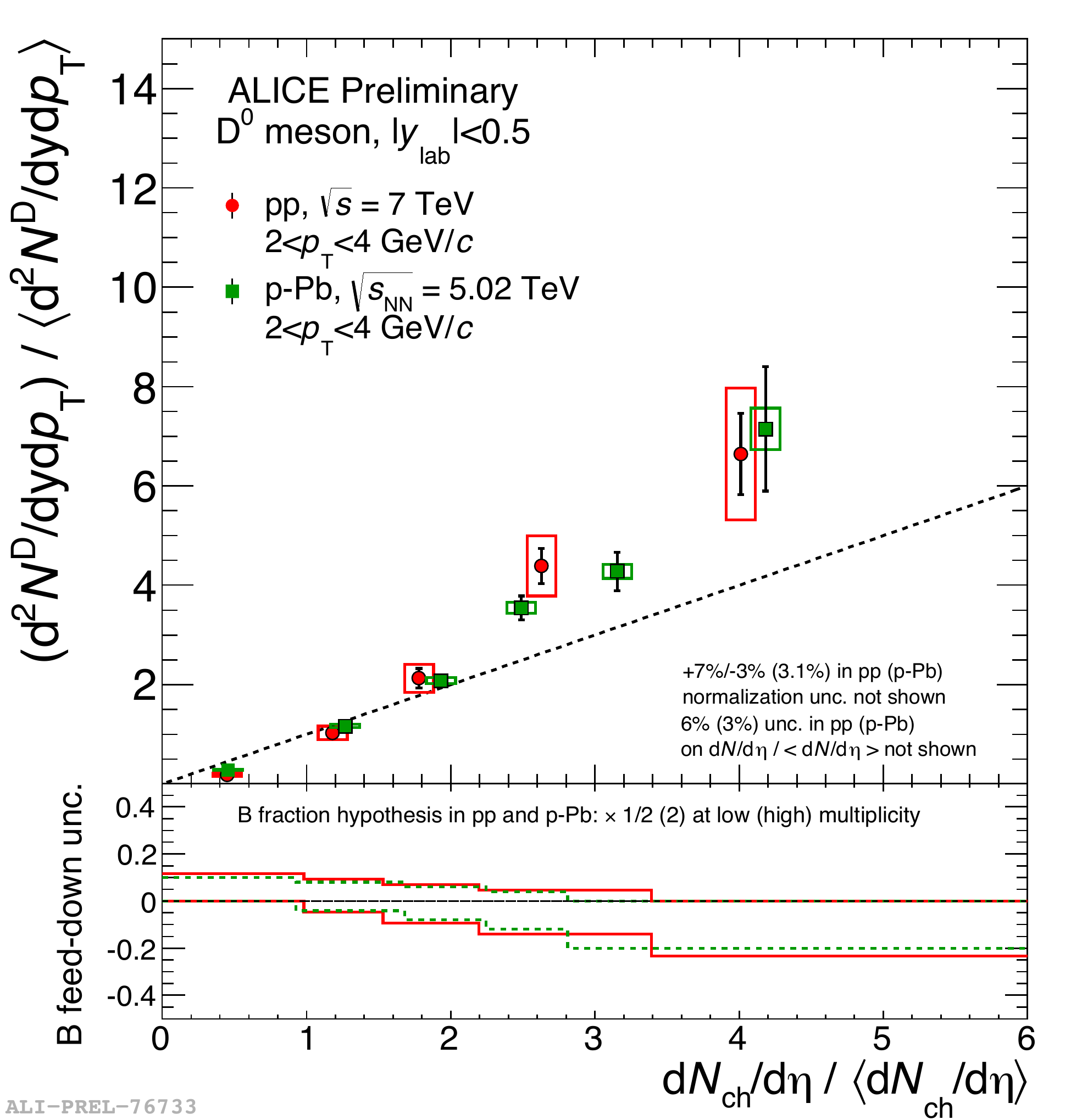}}
 \hspace{2mm}
 \subfigure
   {\includegraphics[width=6.5cm]{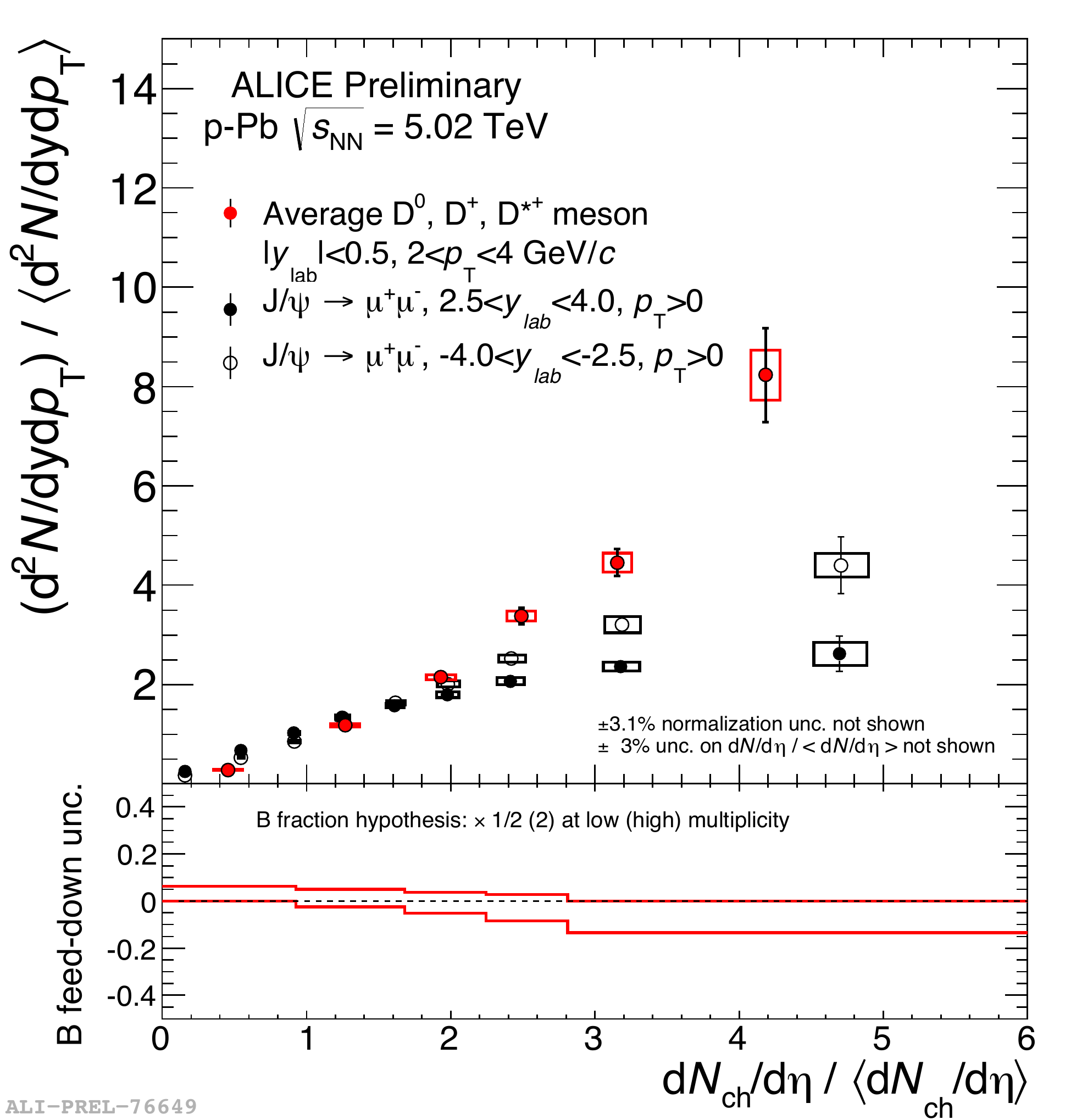}}
\vspace{-0.2cm}
 \caption{Left: self-normalized $\rm D^{0}$ yields for 2 $<p_{\rm T}<$ 4 GeV/$c$  as a function of multiplicity for pp and p--Pb collisions. Right: average $\rm D^{0}$, $\rm D^{+}$ and $\rm D^{*+}$ self-normalized yields compared
to J/$\psi $  in the rapidity intervals  2.5$<y_{\rm lab}<$4.0 and $-4.0 <y_{\rm lab}<-2.5$  for p--Pb collisions}
 \end{figure}

\vspace{-1.cm}
\section{Results}
\vspace{-0.2cm}
The self-normalized yields have been measured for prompt $\rm D^{0}$, $\rm D^{+}$ and $\rm D^{*+}$ mesons. They  are shown in charged-particle multiplicity (dN$_{\rm ch}$/d$\eta$) intervals (Fig.1), since  a simulation study has shown that $N_{\rm tracklets}$/$\langle N_{\rm tracklets} \rangle$ equals
$\rm d \textit{N}_{\rm ch}/ \rm d \eta / \langle d\textit{N}_{\rm ch} / d \eta\rangle$. For all the $\rm D$-meson species, the yield increases  with charged-particle multiplicity. No  $p_{\rm T}$ dependence of this trend  has been observed. 

The left panel of Fig.1 shows $\rm D^{0}$  self-normalized yields for pp and p--Pb collisions. Both systems show an increase of the yield with  charged-particle multiplicity. 
The  trend for pp collisions can be interpreted as being due
to strong hadronic activity connected with charm production and to the
presence of MPIs affecting the hard momentum scale relevant for heavy-quark
production.  In the p--Pb case, it should be considered that high-multiplicity events can also originate from a higher number of nucleon-nucleon collisions in the nuclear interaction. 
The right panel of Fig.1 shows the average values of  self-normalized yields for  $\rm D^{0}$, $\rm D^{+}$ and $\rm D^{*+}$  and  J/$\psi $ in p--Pb collisions.  J/$\psi $  yields have been measured in 2.5$<y_{\rm lab}<$4.0 (p-going direction) and $-4.0 <y_{\rm lab}<-2.5$ (Pb-going direction) and they show an increase with charged-particle multiplicity.
However a quantitative comparison of  J/$\psi $ and $\rm D$-meson yields has to take into account  the rapidity dependence of  cold
 nuclear matter effects such as gluon shadowing, that depends on the Bjorken $x$ value of the parton involved in the process producing charm, and energy loss in cold nuclear matter \cite{Javier}.

The average value of $Q_{\rm pPb}$ for prompt $\rm D^{0}$ and  $\rm D^{*+}$ mesons is shown in the left panels of Figs. 2 and 3  for the V0A and  ZNA estimators, respectively. A bias can be observed in the V0A measurement,
where the low-multiplicity $Q_{\rm pPb}$  is below unity in all  six $p_{\rm T}$ bins, while the ZNA  measurement is compatible with unity within systematic and statistical uncertainties at
both low and high event activity. These $Q_{\rm pPb}$ results are compared  with the  ones obtained for charged particles,  shown in the right panels of Figs. 2 and 3.
These comparisons demonstrate that the $Q_{\rm pPb}$ of high $p_{\rm T}$ ($>$ 8 GeV/$c$) charged particles feature a similar pattern  as the one of $\rm D$ mesons, confirming the presence of a bias in the
V0A-based determination of $\langle N_{\rm coll}^{\rm Glauber}\rangle$  that is reduced using the ZNA estimator. This indicates that  the determination of 
$\langle N_{\rm coll}\rangle$ depends on the rapidity region in which the event activity measurement is performed. With the least biased estimator (ZNA) we observe $Q_{\rm pPb}$ being compatible with unity for all
multiplicites and $p_{\rm T}$.
Details on the $N_{\rm coll}$ bias have been presented at this conference and they are described in \cite{Alberica}.

In conclusion, the D-meson self-normalized yields show an increasing trend with increasing charged-particle multiplicity. The trends observed  for pp and p--Pb collisions are compatible within uncertaintes. 
The $Q_{\rm pPb}$ results for $\rm D$ mesons are qualitatively similar to the ones obtained for high $p_{\rm T}$ charged particles. 
In particular the ZNA  measurement shows  no multiplicity dependence of the $\rm D$-meson production in p--Pb collisions relative
 to binary scaling of pp production cross sections while  V0A  results show a similar  bias as observed  for high-$p_{\rm T}$ charged particles.

\begin{figure}[h]
  \centering
 \subfigure
   {\includegraphics[width=6.3cm]{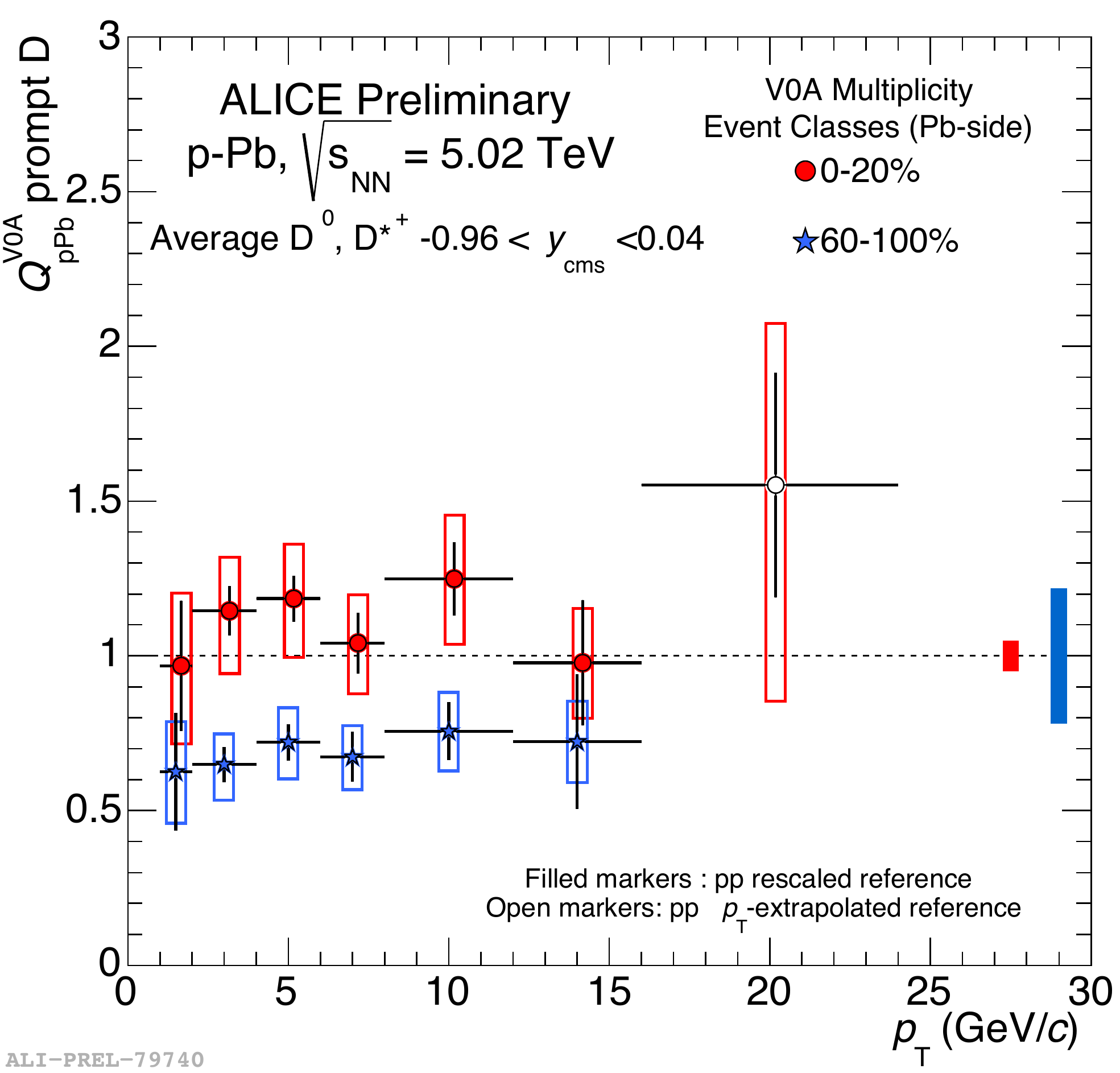}}
 \hspace{2mm}
 \subfigure
   {\includegraphics[width=6.5cm]{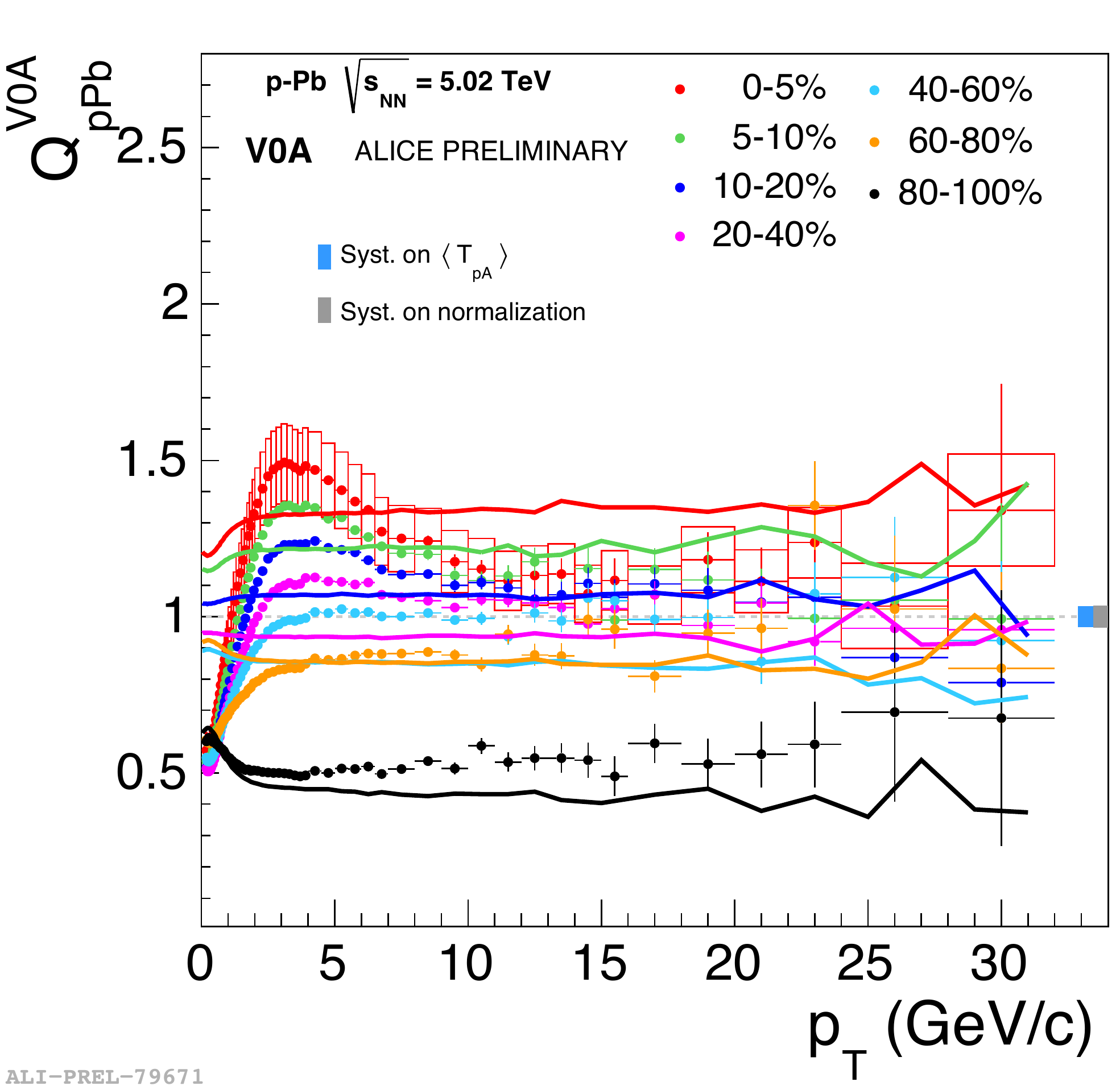}}
 \caption{Left: average $p_{\rm T}$ differential $Q_{\rm pPb}$ of $\rm D^{0}$ and  $\rm D^{*+}$ mesons in the 0-20\% and 60-100\% V0A event activity classes. Right: $p_{\rm T}$ differential $Q_{\rm pPb}$ of
charged hadrons  in seven V0A event activity classes.}

 \centering
 \subfigure
   {\includegraphics[width=6.3cm]{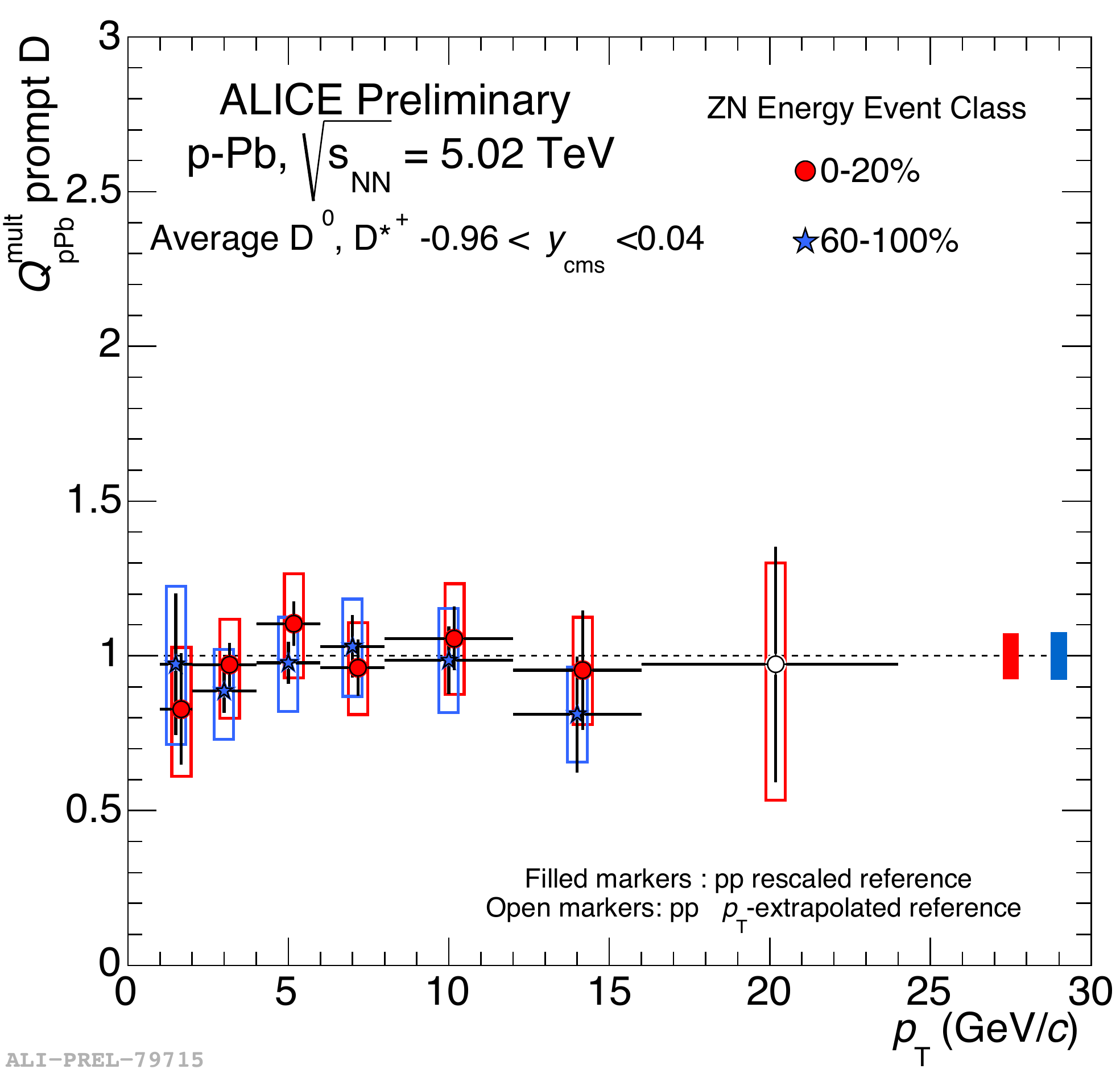}}
 \hspace{2mm}
 \subfigure
   {\includegraphics[width=6.3cm]{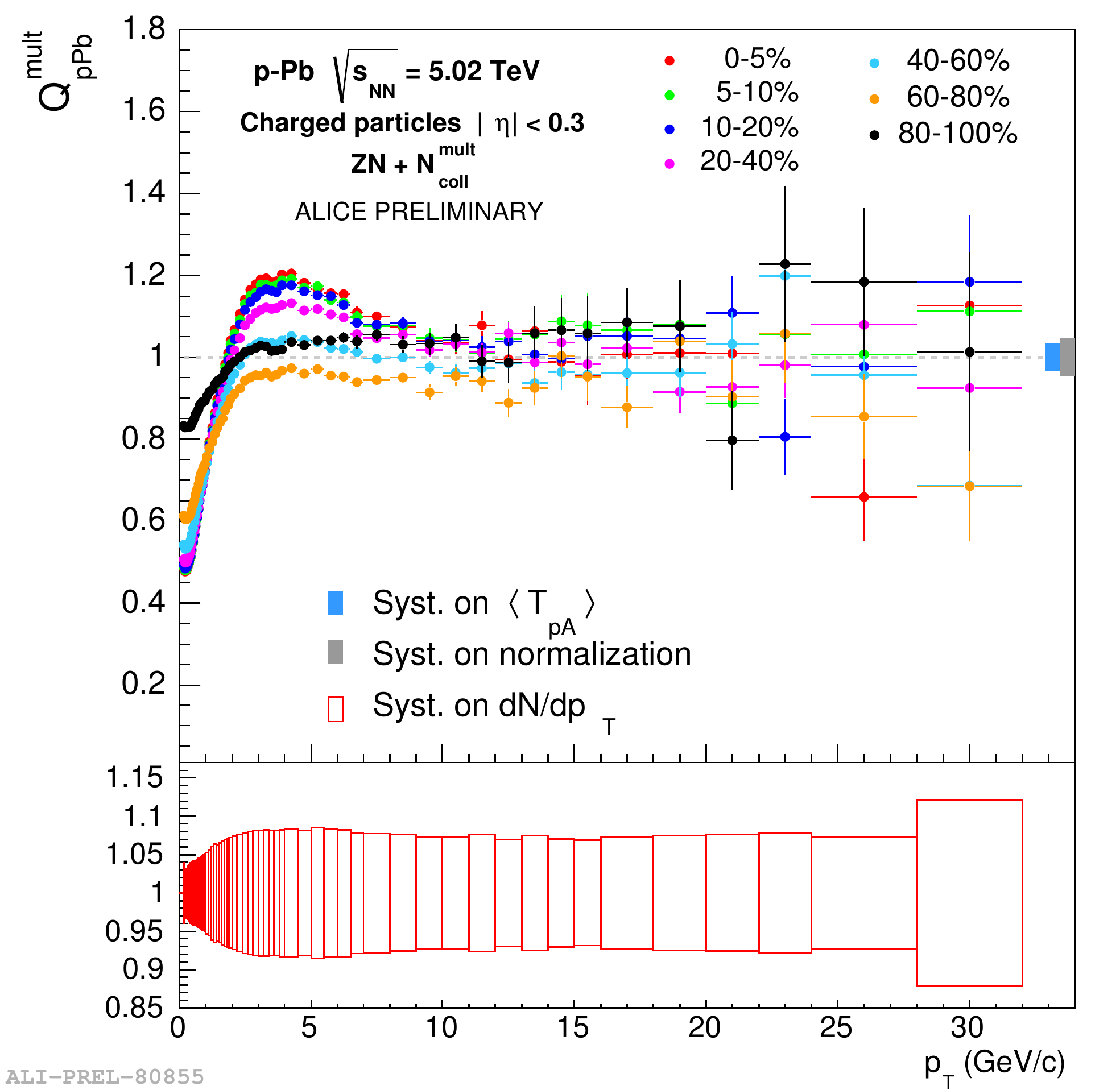}}
 \caption{Left: average $p_{\rm T}$ differential $Q_{\rm pPb}$ of $\rm D^{0}$ and  $\rm D^{*+}$mesons in the 0-20\% and 60-100\% ZNA event activity classes. Right: $p_{\rm T}$ differential $Q_{\rm pPb}$ of
charged hadrons  in seven ZNA event activity classes.}
 \end{figure}

%% The Appendices part is started with the command \appendix;
%% appendix sections are then done as normal sections
%% \appendix

%% \section{}
%% \label{}

%% References
%%
%% Following citation commands can be used in the body text:
%% Usage of \cite is as follows:
%%   \cite{key}         ==>>  [#]
%%   \cite[chap. 2]{key} ==>> [#, chap. 2]
%%

%% References with BibTeX database:

%\bibliographystyle{elsarticle-num}
%\bibliography{<your-bib-database>}

%% Authors are advised to use a BibTeX database file for their reference list.
%% The provided style file elsarticle-num.bst formats references in the required Procedia style

%% For references without a BibTeX database:

\end{document}